\documentclass[a4paper,12pt]{JHEP3}
\usepackage{epsfig}

\title{The critical dimension of bosonic string theory in AdS space-time}
\author{Ian Davies \\ Centre for Particle Theory, University of
Durham, Durham DH1 3LE, UK \\ Email: I.J.Davies@durham.ac.uk}
\author{Paul Mansfield \\Centre for Particle Theory, University of
Durham, Durham DH1 3LE, UK \\ Email: P.R.W.Mansfield@durham.ac.uk}
\abstract{The Polyakov bosonic string is quantised in Euclidean Anti-de Sitter
space-time using functional methods. Regularisation of the functional
determinants using both heat-kernel and zeta function techniques shows
that the $AdS_{D+1}$ string reduces to the Liouville field theory, as in the
flat space-time case. A Feynman graph expansion is used to evaluate
(approximately) the coefficient multiplying the Liouville action; this
then gives the critical dimension for this space-time as 22 (that is, 21 flat
directions plus one radial direction).}


\preprint{hep-th/0108212}

\begin{document}

\newcommand{\x}{{\bf x}}
\newcommand{\y}{{\bf y}}
\newcommand{\z}{{\bf z}}
\newcommand{\bp}{{\bf p}}
\newcommand{\A}{{\bf A}}
\newcommand{\B}{{\bf B}}
\newcommand{\p}{\varphi}
\newcommand{\del}{\nabla}
\newcommand{\be}{\begin{equation}}
\newcommand{\ee}{\end{equation}}
\newcommand{\bq}{\begin{eqnarray}}
\newcommand{\eq}{\end{eqnarray}}
\newcommand{\ba}{\begin{eqnarray}}
\newcommand{\ea}{\end{eqnarray}}
\def\r{\nonumber\cr}
\def\hf{\textstyle{1\over2}}
\def\qr{\textstyle{1\over4}}
\def\Sc{Schr\"odinger\,}
\def\sc{Schr\"odinger\,}
\def\'{^\prime}
\def\>{\rangle}
\def\<{\langle}
\def\-{\rightarrow}
\def\dbd{\partial\over\partial}
\def\tr{{\rm tr}}
\def\pd{\partial}

\section{Introduction}

It has long been suspected that there should exist a string-theoretic
description of gauge theory. This idea has it's origins back in the
work of K. Wilson ~\cite{Wilson}, where he showed that confinement in
QCD could be understood within lattice gauge theory as the binding
together of quarks on the lattice by colour-electric flux
tubes. Unfortunately, this does not provide us with a microscopic
description of the confinement mechanism in QCD, since this result is
valid only in the strong coupling expansion and has no continuum
limit. The large $N$ expansion of QCD discovered by 't
Hooft~\cite{'tHooft} showed that the large-$N$ limit of gauge theories
has a natural description in terms of topologically planar Feynman
diagrams; this is consistent with a string-theoretic approach in terms
of worldsheets. Building on these ideas, Polyakov~\cite{55} made the conjecture that
there ought to be a continuum string theory description of certain
gauge invariant objects in QCD, namely the Wilson loops. More
recently, the $AdS$/CFT correspondence~\cite{21}~\cite{24}~\cite{49}
has provided us with a string-theoretic description of a 4-dimensional
conformal field theory with $\mathcal{N}=4$ supersymmetry. This
conjecture has been checked and confirmed in numerous papers; however, we
do not yet have a concrete worldsheet realization of the
correspondence. The search for such a string-theoretic formulation of QCD forms the basis for the present work. In the long term,
it is to be hoped that such an understanding will lead towards a microscopic
understanding of strongly coupled QCD. In the meantime, we restrict
ourselves to a more modest aim - to identify and develop the
calculational techniques that may help derive the $AdS$/CFT
correspondence (or some similar duality) from ``first principles''.

A promising place to begin looking for a worldsheet understanding of
the correspondence is the work of
Polyakov~\cite{55}~\cite{54}~\cite{58}. The basic idea is to
reformulate pure gauge theory entirely in terms of objects called {\it
loop functionals}, an example of which is the Wilson loop. All
equations of motion, etc., are to be rewritten in terms of functionals
and operations which are defined on {\it loop space} - that is, the
space of arbitrary continuous closed loops. The resulting equations of
motion (known as the {\it loop equations}~\cite{Makeenko}) are then
conjectured to be soluble in analogy with the Feynman integrals over
trajectories in standard field theory. Instead of a sum over paths
whose end points are fixed, we sum over {\it surfaces} whose
boundaries trace out the contour on which the Wilson loop is
defined. Such a sum over surfaces is precisely what we encounter in
string theory. The aim, then, is to choose the string theory background
in such a way that we reproduce the gauge theory equations of motion,
via the loop equation. This approach to gauge fields -- strings
duality is much more general than $AdS$/CFT - the specific string
background which we choose to use is determined by the properties of
whatever gauge theory we have decided to study. In fact, Anti-de
Sitter space-time is a simple example of a background which possesses
{\it zigzag symmetry} (a property of the Wilson loop) at certain
points. The claim, then, is that we should place the ends of an open
string at these zigzag-symmetric points in $AdS_{D+1}$; the
ends of the string then trace out the contour of a Wilson loop in a 
$D$-dimensional gauge theory. In this way, we arrive at a dual
description of the $D$-dimensional Wilson loop in terms of a
$(D+1)$-dimensional string theory in $AdS$ space-time. This
construction does not use D-branes at any stage, and as such may be a
way of developing an understanding of gauge fields -- strings dualities for
non-supersymmetric gauge theories.

It has been shown~\cite{58} that the classical minimal area in
Euclidean $AdS$ space
approximately satisfies the loop equations for non-supersymmetric $SU(N)$ gauge
theory in the large-$N$ limit. With this in mind, it is of interest to
study the quantum theory of string worldsheets in $AdS$. By using a
sensible set of coordinates, we can write the Euclidean version of the
$AdS_{D+1}$ metric (which is the Lobachevsky space on the upper-half plane) in a
form which is conformal to $\mathbf{R}^{D+1}$:
\be
ds^{2} =
\frac{R^{2}}{y^{2}}\left(dy^{2}+\sum_{i=1}^{D}dX_{i}^{2}\right) \label{met}
\ee
where $R$ is the radius of the $AdS$ space-time. In this paper we
attempt to quantise the Polyakov bosonic string in
this background. We begin by re-writing the string partition function
in a new set of $(D+1)$ variables (section 2.1) and consider the quantisation of
$D$ of them using zeta function regularisation techniques (section
2.2). The string action is quadratic in these $D$ fields, which correspond to the $X_{i}$'s in the metric
(\ref{met}). We initially work
in the flat gauge on the worldsheet, $g_{ab}=\delta_{ab}$, and then in
section 2.3 we calculate the extra terms introduced by working in the conformal gauge
$g_{ab}=e^{\phi(\xi)}\delta_{ab}$ (the presence of the ``Liouville
mode'' $\phi(\xi)$ leads to a conformal anomaly, as in the flat
space-time case). Having done this, we analyse the symmetries of the
resultant effective action; this then allows us to construct a Feynman
integral expansion to integrate out (approximately) the remaining
space-time field (section 2.4). This computation yields an approximate value for
the critical dimension for bosonic strings in $AdS$ of 22. A
discussion of this result is included in section 3. 

\section{Quantisation of the bosonic string in $AdS$}

\subsection{Change of variables}

We begin by considering the usual expression for the partition
function of a bosonic string in a $(D+1)$-dimensional $AdS$ space-time
with Euclidean signature. This is a non-linear sigma model, with target
space metric (\ref{met}):
\be
Z=\int \mathcal{D}g\mathcal{D}X\mathcal{D}y e^{-S}
\ee
where
\be
S = \frac{R^{2}}{4\pi}\int
d^{2}\xi\sqrt{g}g^{ab}\frac{\partial_{a}X^{i}\partial_{b}X^{i}}{y^{2}}+\frac{R^{2}}{4\pi}\int
d^{2}\xi\sqrt{g}g^{ab}\frac{\partial_{a}y\partial_{b}y}{y^{2}} \label{action}
\ee
Here, $i=1,..,D$ and we have set $\alpha^{\'}=1$. We do not consider any other background fields (such as the
antisymmetric tensor field) here; we are only concerned with the
sector of the string theory that has already been shown to satisfy the
loop equations classically~\cite{54}. Functional quantisation of
string theory relies on setting up all the integrals in a manifestly
reparametrisation invariant way. Weyl invariance, which is a symmetry
of the classical theory, is then broken quantum-mechanically through
regularisation of the functional integrals. We therefore need to check
by hand the conditions under which Weyl invariance is restored at the
quantum level; as is explained below, this is what leads to the
existence of a critical dimension in this formalism.

In order to use functional integrals to quantise the theory, we need a
proper definition of the measure. Following~\cite{13}, we construct
the volume element in analogy with the construction of such elements
in finite-dimensional integrals. This requires a definition of an
inner product of worldsheet fields, and this definition must by
construction respect
reparametrisation invariance. The target space metric (\ref{met}) thus
dictates that the inner product on variations of the $X$-fields is
\be
(\delta X^{i}_{1},\delta X^{i}_{2}) \equiv \int d^{2}\xi
\frac{\sqrt{g}}{y^{2}}\delta X^{i}_{1}\delta X^{i}_{2} \label{ip}
\ee
We see that the functional integration over the
$X$-fields is a simple Gaussian, and gives
\be
\mbox{Det}^{-\frac{D}{2}}\left[-\frac{1}{\mu^{2}}\frac{y^{2}}{\sqrt{g}}\partial_{a}\left(\frac{1}{y^{2}}\sqrt{g}g^{ab}\partial_{b}\right)\right]\equiv
\mbox{Det}^{-\frac{D}{2}}\left(\frac{\Omega}{\mu^{2}}\right) \label{det}
\ee
In order to proceed with the quantisation we need to
regularise this infinite dimensional functional determinant, which
should be thought of as an infinite product of the eigenvalues of
$\Omega / \mu^{2}$. Hence, we have introduced the mass scale $\mu^{2}$ in order to make the determinant
dimensionless. However,
it is convenient to rewrite the partition function in a new set of
variables,  
\be
W^{i}(\xi) = \frac{X^{i}(\xi)}{y(\xi)}, {\hskip 1cm} y(\xi) = e^{\chi(\xi)}
\ee
and then define a new field $f(\xi)$ to be
\be
f(\xi) \equiv g^{ab}\pd_{a}\chi\pd_{b}\chi+\Delta\chi
\ee
where $\Delta$ is the usual covariant Laplacian
operator associated with the
worldsheet metric $g_{ab}$. This puts the partition function (before
integrating out any of the fields) into the
form
\be
Z =
\int\mathcal{D}g\mathcal{D}W\mathcal{D}f\left[\mbox{Det}^{-1}\left(\frac{\Delta +
f}{\mu^{2}}\right) \right]e^{-S^{\'}} \label{sig}
\ee
with
\be
S^{\'} = \frac{R^{2}}{4\pi}\int d^{2}\xi\sqrt{g}\left(f
+ W\cdot (\Delta + f)\cdot W\right)
\ee
The functional determinant appearing here is the Jacobian associated
with this change of variables. As described above, we must define the
measure associated with the $W$-fields in a reparametrisation
invariant way; this forces us to use the inner product
 $$(\delta W^{i}_{1},\delta W^{i}_{2}) \equiv \int d^{2}\xi
\sqrt{g}\:\delta W^{i}_{1}\delta W^{i}_{2},$$
and so the functional integration over the
$W$-field is now itself a straightforward Gaussian. This integral gives another factor
of the Jacobian determinant given above; $W$ is a $D$-dimensional
vector of scalar fields, and so the result of carrying out the $W$-integration is just
\be
Z = \int
\mathcal{D}g\mathcal{D}f\left[\mbox{Det}^{-\frac{D+2}{2}}\left(\frac{\Delta +
f}{\mu^{2}}\right)\right]\exp \left(-\frac{R^{2}}{4\pi}\int d^{2}\xi\sqrt{g}f\right) \label{pf}
\ee

\subsection{Regularisation of the determinant (flat worldsheet)}

Our goal now is to regulate the infinite-dimensional determinant
arising in the expression (\ref{pf}). To do this, we will begin by assuming that
the string worldsheet is flat in order to extract the
$\chi$-dependence of the determinant. In this case, $\Delta + f \-
-\pd_{a}^{2} + (\pd_{a}\chi)^{2}-\pd_{a}^{2}\chi$ and we can write
down a {\it heat-kernel equation} for this operator:
\be
(-\pd_{a}^{2} + f)K(\xi,\xi^{\'};t) = -\mu^{2}\frac{\pd}{\pd
t}K(\xi,\xi^{\'};t) \label{heateq}
\ee
with the initial condition $$\lim_{t \- 0}K(\xi,\xi^{\'};t) =
\delta^{2}(\xi-\xi^{\'})$$ We can write the logarithm of the
determinant we are interested in by using the zeta function, defined by:
$$
\zeta (s) = \sum_{\lambda}\lambda^{-s}
$$
where $\lambda$ are the eigenvalues of $-\pd_{a}^{2}+f$, and so
$$\ln \mbox{Det}\left(\frac{-\pd_{a}^{2}+f}{\mu^{2}}\right) = -\zeta^{\'}(0)$$ We can represent the zeta
function as an integral over the trace of $K(\xi,\xi^{\'};t)$: $$\zeta (s) = \frac{1}{\Gamma (s)}\int_{0}^{\infty}dt\left(t^{s-1}\mbox{Tr}K\right)$$
Hence, the regularisation
procedure relies on knowing the heat-kernel $K(\xi,\xi^{\'};t)$ for
$\Omega$. In order to do this, we have to make an approximation. To
begin with, we can
work to the leading terms in a derivative expansion of $\chi$, so that
$(\pd_{a}\chi)$ is constant. In this case, $f$ itself is a
positive-definite constant and as such will act as a mass term in the
heat equation (\ref{heateq}). We can now write down $K(\xi,\xi^{\'};t)$ explicitly:
\be
K(\xi,\xi^{\'};t) = \frac{\mu^{2}}{4\pi t}e^{-\frac{\mu^{2}|\xi-\xi^{\'}|^{2}}{4t}-\frac{ft}{\mu^{2}}}
\ee
Substituting this heat-kernel into the expression for the zeta
function then gives $$\zeta (s) = \int d^{2}\xi
\left[\mu^{2}\left(\frac{f}{\mu^{2}}\right)^{1-s}\frac{1}{4\pi (s-1)}\right],$$
which then implies that
\be\ln \mbox{Det}\left(\frac{-\pd_{a}^{2}+f}{\mu^{2}}\right) = \frac{1}{4\pi}\int
d^{2}\xi \left(f-f\ln\left(\frac{f}{\mu^{2}}\right)\right)\ \label{logs}
\ee
We can now substitute this determinant back into our expression for
the partition function (\ref{pf}). At present, our result appears to
depend explicitly on the arbitrary mass scale $\mu^{2}$; however, this
dependence can be absorbed by a redefinition of the radius of our
$AdS$ space-time, $R$ \footnote{It is perhaps worth mentioning that the above renormalisation property may
have implications for $AdS$/CFT. If we naively apply the relation
$R^{2}=g_{YM}^{2}N$ to the above redefinition of $R$, we arrive at an
expression for the dual gauge theory coupling constant as a function
of the scale $\Lambda$. In other words, we get a gauge theory
$\beta$-function $$\beta \equiv
\Lambda\frac{dg_{YM}(\Lambda)}{d\Lambda} = \frac{(D+2)}{\sqrt{N}}$$ It is not clear
what this quantity really means. Obviously we haven't even quantised
all the space-time fields at this stage (and those that we have, have
been calculated only approximately), so the fact that $\beta \neq
0$ should not worry us unduly. Also, the construction of
Polyakov~\cite{55} is based on just a pure Yang-Mills theory, not a
supersymmetric one such as $\mathcal{N}=4$ super-Yang Mills. But it is
intriguing that one side of the $AdS$/CFT relationship
$R^{2}=g_{YM}^{2}N$ appears to be determined by worldsheet
renormalizability in this calculation.}:
$$
R^{2} = \frac{(D+2)}{2}\ln\left(\frac{\Lambda^{2}}{\mu^{2}}\right)
$$
($\Lambda$ is a renormalisation scale, in the same sense as
$\Lambda_{QCD}$). So, we arrive at the following expression for the
partition function, having integrated out the $W$-fields within the
approximation that $f$ is constant and using a flat worldsheet metric:
\be
Z = \int \mathcal{D}f\exp\left(-\frac{(D+2)}{8\pi}\int
d^{2}\xi\left(f-f\ln\left(\frac{f}{\Lambda^{2}}\right)\right)\right) \label{part}
\ee

\subsection{The Weyl anomaly}

An important aspect of string physics is the Weyl anomaly - in fact,
in this functional formalism it is the organising
principle. Reparametrisation invariance is present from the outset,
but it is the restoration of Weyl invariance by hand at the quantum level that
determines the consistency conditions, such as the existence of a critical
dimension. For example, in flat space-time it is only in
26 dimensions that the Faddeev-Popov determinant arising
from fixing the conformal gauge $g_{ab}=e^{\phi}\delta_{ab}$ is
exactly cancelled by the functional determinant obtained by quantising
the space-time fields $X^{\mu}$~\cite{13}. Away from this critical dimension we
are left with the {\it Liouville field theory}, which in the conformal
gauge is
\be
\int \mathcal{D}\phi\exp\left(-\int
d^{2}\xi\left[(\pd_{a}\phi)^{2}+\lambda e^{\phi}\right]\right)
\ee
($\lambda e^{\phi}$ is a finite contribution left over from the addition of a
counterterm required for renormalizability). Since the functional
measure $\mathcal{D}\phi$ depends on $\phi$ itself in a highly
complicated way, this integral is intractable. The problem of solving
Liouville field theory is thus central to the understanding of string
theories away from their critical dimensions, since if we were able to
correctly integrate out the Liouville mode, the resulting partition
function would be Weyl invariant in all space-time dimensions.

Do we find the same kind of structure in $AdS$ space-time? We need to
determine the explicit $\phi$-dependence of the determinant
$$\mbox{Det}(\Delta + f)$$
in order to begin to answer this question. There are several different ways of
doing this; here we will use the heat-kernel expansion method used
in~\cite{13}. Clearly, the determinant given in (\ref{det}) is equivalent to $\mbox{Det}(\Delta
+ f)$, so we will compute the variation of $\ln \mbox{Det}\Omega$ with
respect to $\phi$. We can then use the result as a consistency
check on the result obtained in the previous section (we should find
agreement in the $f$-coefficient between this calculation
and the result of replacing $\mu^{2}\rightarrow \mu^{2}e^{\phi}$ in
(\ref{logs}) and varying it with respect to $\phi$,
since both $\mu$ and $\phi$ always appear in the combination $\mu^{2}e^{\phi}$). In the conformal gauge, then, we are
interested in
$$
\mbox{Det}\left[-e^{-\phi}\pd_{a}^{2}+2e^{-\phi}(\pd_{a}\chi)\pd_{a}\right]
$$
(we have dropped the $\mu$-dependence, since we have already calculated it) and we must now use the inner product (\ref{ip}) for worldsheet scalar
fields. In analogy with the identity
$$\delta\ln \mbox{det} M =
\int_{0}^{\infty}dt \:\mbox{tr}\left(\delta M e^{-tM}\right)
$$
which holds for any finite-dimensional matrix $M$ with
positive-definite eigenvalues, we can define the regularised
determinant of $\Omega$ by
\be
\delta_{\phi}\ln \mbox{Det}\Omega =
\int_{\epsilon}^{\infty}dt\:\mbox{Tr}\left(\delta_{\phi}\Omega
e^{-t\Omega}\right) \label{rep}
\ee
where $\epsilon$ is a short-time cutoff (we can do this, since
$\Omega$ is hermitian with respect to the inner product (\ref{ip})). Performing the $t$-integral, and noting that
$\delta_{\phi}\Omega = -\delta\phi\Omega$, we obtain
\be
\delta_{\phi}\ln \mbox{Det}\Omega = -\mbox{Tr}\left(\delta\phi(\xi)e^{-\epsilon\Omega}\right)
\ee
Again we make use of a
heat-kernel equation, this time for $\Omega$. We can represent
$e^{-t\Omega}$ as
$$
e^{-t\Omega}f(\xi) = \int d^{2}\xi^{'}e^{\phi(\xi^{\'})-2\chi(\xi^{\'})}K(\xi,\xi^{'};t)f(\xi^{'})
$$
and hence
$$
\frac{\partial K(\xi,\xi^{\prime};t)}{\partial t} = -\Omega K(\xi,\xi^{\prime};t)
$$
with the initial condition
$$
\lim_{t\rightarrow 0}K = e^{2\chi(\xi^{\'})-\phi(\xi^{'})}\delta^{2}(\xi-\xi^{'})
$$

The fact that we have introduced a short-time cutoff means that we are
only interested in {\it local} worldsheet effects. Therefore, we can
suppose that the worldsheet metric we are considering is a
perturbation about a flat metric, and we should be able to evaluate
the heat-kernel for $\Omega$ by expanding around the heat-kernel for
the flat-space Laplacian $\Delta$. The ``unperturbed'' heat-kernel satisfying
the heat equation for $\Delta$ is just
\be
K_{0} = \frac{e^{2\chi(\xi^{\prime})-\phi(\xi^{\prime})}}{4\pi
t}e^{-\frac{\vert \xi-\xi^{'}\vert^{2}}{4t}} \label{kern1}
\ee
We know that
$$
\int d^{2}\xi^{'}e^{\phi(\xi^{\prime})-2\chi(\xi^{\prime})}K_{0}(\xi,\xi^{'};t-t^{'})K_{0}(\xi^{'},\xi;t^{'}) = K_{0}(\xi,\xi;t) = \frac{e^{2\chi(\xi)-\phi(\xi)}}{4\pi t}
$$
from looking at the expansion of the heat-kernel in eigenfunctions of
its operator. Remember that the Weyl scaling behaviour of $\Omega$ is
given in terms of the {\it trace} of it's heat-kernel, so we can use
the following {\it heat-kernel expansion} to evaluate $K(\xi,\xi;\epsilon)$:
\begin{eqnarray}
K^{t}(\xi,\xi) &=& \frac{1}{4\pi t} + \int_{0}^{t}dt^{\prime}\!\int d^{2}\xi^{'}e^{2\chi-\phi}K^{t-t^{\prime}}_{\xi\xi^{'}}V(\xi^{'})K^{t^{\prime}}_{\xi^{'}\xi} \nonumber \\
&+& \int_{0}^{t}dt^{\prime}\!\!\int_{0}^{t^{\prime}}dt^{\prime\prime}\!\!\int d^{2}\xi^{'}e^{2\chi-\phi}\int d^{2}\xi^{''}e^{2\chi-\phi}K^{t-t^{\prime}}_{\xi\xi^{'}}V(\xi^{'})K^{t^{\prime}-t^{\prime\prime}}_{\xi^{'}\xi^{''}}V(\xi^{''})K^{t^{\prime\prime}}_{\xi^{''}\xi} \nonumber \\
&+& \cdots \label{series}
\end{eqnarray}
where we have expanded the operator $\Omega$ around the flat-space
Laplacian
$$
\Omega = \Delta + V(\xi)
$$ 
and used the heat-kernel (\ref{kern1}). We need to calculate all the terms in (\ref{series}) which are of
$O(1)$ in the time variable $t$. Terms of order $t$ will not
contribute when we send the cutoff $\epsilon$ to zero. The heat
equation for $K_{0}$ implies that the heat-kernel for the operator
$-e^{-\phi(\bar{\xi})}\partial_{a}^{2}$ where $e^{-\phi(\bar{\xi})}$
is a constant must be
\begin{equation}
\bar{K}_{0}(\xi,\xi^{\prime};t) = \frac{e^{2\chi(\xi^{\prime})-\phi(\xi^{\prime})}}{4\pi
t}e^{\phi(\bar{\xi})}e^{\frac{-\vert
\xi-\xi^{\prime}\vert^{2}}{4t}e^{\phi(\bar{\xi})}} \label{hk}
\end{equation}
and so we can write $\Omega$ as
$$
\Omega = e^{-\phi(\bar{\xi})}\Delta - V(\xi)
$$
where
\begin{eqnarray}
V(\xi^{'}) &=&
\left((\xi^{'}-\bar{\xi})^{a}\partial_{a}\phi-\frac{1}{2}(\xi^{'}-\bar{\xi})^{a}(\xi^{'}-\bar{\xi})^{b}\left(\partial_{a}\phi\partial_{b}\phi-\partial_{a}\partial_{b}\phi\right)\right)e^{-\phi(\bar{\xi})}\Delta^{'}
\nonumber \\ &-&
2\left(1-(\xi^{'}-\bar{\xi})^{a}\partial_{a}\phi+\frac{1}{2}(\xi^{'}-\bar{\xi})^{a}(\xi^{'}-\bar{\xi})^{b}\left(\partial_{a}\phi\partial_{b}\phi-\partial_{a}\partial_{b}\phi\right)\right)
\nonumber \\
&\times&\left(\partial_{a}\chi+(\xi^{'}-\bar{\xi})^{b}\partial_{b}\partial_{a}\chi\right)e^{-\phi(\bar{\xi})}\partial_{a}^{'} \nonumber
\end{eqnarray}
We are now faced with the task of substituting this expression for $V$
into the heat-kernel expansion (\ref{series}) and evaluating all the
relevant terms, using the heat-kernel (\ref{hk}). This calculation,
although now conceptually straightforward, is rather lengthy and
technical. Essentially, it involves calculating a series of
2-dimensional Gaussian integrals; this is most easily done by
constructing a ``generating function'' and taking derivatives in order
to generate the required terms. For the reader's convenience the
details are given in Appendix A; here, we simply quote the result for the trace of the heat-kernel of $\Omega$:
$$
K(\xi,\xi;\epsilon)= e^{2\chi(\xi)-\phi(\xi)}\left[\frac{e^{\phi}}{4\pi\epsilon}-\frac{1}{24\pi}\partial_{a}^{2}\phi+\frac{1}{4\pi}\left(\partial_{a}^{2}\chi-(\partial_{a}\chi)^{2}\right)+O(\epsilon)\right]
$$
We can now use this result to evaluate $\delta_{\phi}\ln \mbox{Det}\Omega$
by taking $\epsilon \rightarrow 0$:
\bq
\delta_{\phi}\ln \mbox{Det}\Omega &=& \int
d^{2}\xi\delta\phi\left[\frac{1}{24\pi}\partial_{a}^{2}\phi-\frac{1}{4\pi}\left(\partial_{a}^{2}\chi-(\partial_{a}\chi)^{2}\right)\right]\nonumber
\\
&=& \int
d^{2}\xi\delta\phi\left[\frac{1}{24\pi}\partial_{a}^{2}\phi+\frac{1}{4\pi}f\right]
\label{Weyl}
\eq
We have removed the divergent term by adding a counterterm to the
original string action. Notice that replacing $\mu^{2}\rightarrow
\mu^{2}e^{\phi}$ in expression (\ref{logs}) and varying it with respect to $\phi$
gives 
$$
\delta_{\phi}\ln \mbox{Det}\left(\frac{-\pd_{a}^{2}+f}{\mu^{2}e^{\phi}}\right) =
\int
d^{2}\xi\left[\frac{1}{4\pi}f\right]\delta\phi,
$$
and we have agreement in the $f$-coefficients in this expression and
(\ref{Weyl}), as expected. It is
important to note that the
result (\ref{Weyl}) is correct for {\it all} $f$ - we have not made any
approximation here. Therefore, we can be sure that this result
captures all of the coupling between the Liouville mode $\phi(\xi)$
and $f(\xi)$.

Combining the result for the Weyl anomaly with the $f$-dependence
derived in (\ref{part}), we arrive at the following expression for the
string partition function in $AdS_{D+1}$:
\be
Z = \int \mathcal{D}\phi\int\mathcal{D}f\exp\left(-S^{\'}\right) \times \mbox{Det}^{1/2}_{FP}
\ee
where
\be
S^{\'} = \frac{(D+2)}{8\pi}\int
d^{2}\xi\:\left(f\left(1-\ln\left(\frac{f}{\Lambda^{2}}\right)+\phi\right)-\frac{1}{12}(\pd_{a}\phi)^{2}+\lambda \label{newact}
e^{\phi}\right)
\ee
The $\lambda e^{\phi}$ term in the action is the finite contribution left over
from the counterterm - the value of $\lambda$ is fixed by some
physical renormalisation condition. We have also introduced the
Faddeev-Popov determinant which arises when we fix the conformal gauge
on the worldsheet. This determinant was evaluated in~\cite{13} for the
flat space-time case; since $g_{ab}$ is a local quantity
on the worldsheet, the Faddeev-Popov determinant remains the same
irrespective of the background space-time in which the string
lives. Hence, we quote the result from~\cite{13} that
$$
\mbox{Det}^{1/2}_{FP} = \exp\left(-\frac{13}{48\pi}\int
d^{2}\xi\:\left[(\pd_{a}\phi)^{2}+\lambda e^{\phi}\right]\right) = \exp\left(-\frac{13}{48\pi}S_{L}\right)
$$
We can therefore combine the $f$-independent terms in (\ref{newact})
with the Liouville action arising from the Faddeev-Popov determinant
to obtain
\be
Z = \int
\mathcal{D}\phi\:e^{\frac{D+2-26}{96\pi}S_{L}}\times
\int\mathcal{D}f\: e^{-\frac{(D+2)}{8\pi}\int
d^{2}\xi\:\left[f\left(1-\ln\left(\frac{f}{\Lambda^{2}}\right)+\phi\right)\right]} \label{zed}
\ee

\subsection{Integrating out the $f$-field}

At first sight, integrating out $f$ in equation (\ref{zed}) looks like
an extremely difficult task. However, we can apply some simple
symmetry arguments to give some clues as to how to proceed. Working
for the moment in complex coordinates ($z=\xi_{1}+i\xi_{2}$,
$\bar{z}=\xi_{1}-i\xi_{2}$) so that
$f=\pd\chi\bar{\pd}\chi-\pd\bar{\pd}\chi$, we see that $Z$ has the following
conformal symmetry:
$$
\delta\phi = \epsilon\partial\phi + \partial\epsilon {\hskip 1cm} \delta\chi = \epsilon\partial\chi
$$
with $\bar{\partial}\epsilon =0$. Once the $f$-field has been
integrated out of $Z$, the result will be invariant under the
$\phi$-transformation alone. Crucially, the Liouville action is the unique local theory without dimensionful couplings that has this symmetry. Since our derivation of
the $\phi$-dependence of $Z$ used only a local cutoff, we know that
the result of integrating out $f$ must also be local, and therefore
must be of the form
$$
Z \sim \int \mathcal{D}\phi\: \exp\left(C\int
d^{2}\xi\left[(\pd_{a}\phi)^{2}+\lambda e^{\phi}\right]\right)
$$
The question to be answered is: what is the overall coefficient $C$
multiplying the Liouville action? This will determine the critical
dimension for this system, since when $C=0$ the partition function
will no longer depend on the scale factor $\phi$ and the quantum
theory will be Weyl invariant. 

As was mentioned above, the calculation of the Weyl anomaly captured
all of the coupling between $f$ and $\phi$. This allows us to try and
integrate out $f$ approximately using a Feynman integral-type
approach. We need to find the $\phi$-dependence of $Z$, which from
equation (\ref{zed}) can be
written as
$$
Z = e^{\frac{D+2-26}{96\pi}S_{L}}\times
\int\mathcal{D}f\: e^{-\frac{(D+2)}{8\pi}\int
d^{2}\xi\:\phi f}\times \mbox{Det}^{-\frac{(D+2)}{2}}(-\pd_{a}^{2}+f)
$$
This then must be equal to
$$
Z = \int\mathcal{D}f\: \mbox{Det}^{-N}(\Delta + f)=\int
\mathcal{D}f\:\exp\left(-N\mbox{Tr}\ln (\Delta + f)\right)
$$
where $N=\frac{(D+2)}{2}$. We can expand the exponent,
$$
\mbox{Tr}\ln (\Delta + f) = \mbox{Tr}\ln \Delta + \sum_{n=1}^{\infty}\frac{(-1)^{n}}{n}\mbox{Tr}\left[\left(\Delta^{-1}f\right)^{n}\right] 
$$ 
so that
$$
Z = \mbox{Det}^{-N}\Delta\:\int \mathcal{D}f\:\exp\left[-N\left(-\mbox{Tr}(\Delta^{-1}f)+\frac{1}{2}\mbox{Tr}(\Delta^{-1}f)^{2}-\cdots\right)\right]
$$
We can now use the knowledge gained from our previous calculations to
unpack this expression a little. The first term in the exponential
($\mbox{Tr}(\Delta^{-1}f)$) is actually the one which
generates the $f-\phi$ coupling that we found above (a standard calculation confirms this, and
checks that the coefficients in both cases are the same). The factor
of $\mbox{Det}^{-N}\Delta$ (along with the Faddeev-Popov determinant,
obviously) generates the Liouville action factor appearing in $Z$,
\be
\exp\left(\frac{D+2-26}{96\pi}S_{L}\right) \label{LA}
\ee
and does not enter into the $f$-integration. Rescaling $f$ such that
$\sqrt{N}f=\bar{f}$, we define $\Psi$ to be the relevant $f$-integral:
\be
\Psi = \int
\mathcal{D}\bar{f}\:\exp\left[-\left(\frac{\sqrt{N}}{4\pi}\int
d^{2}\xi\:
\bar{f}\phi+\frac{1}{2}\mbox{Tr}(\Delta^{-1}\bar{f})^{2}-\frac{1}{3\sqrt{N}}\mbox{Tr}(\Delta^{-1}\bar{f})^{3}+\cdots\right)\right] \label{integral}
\ee
(we will omit the bars from $f$ from now on). We know by symmetry arguments that we are looking only for
terms involving $\phi$ of the form $\phi\pd_{a}^{2}\phi$, since only
these are present in the Liouville action (cosmological constant-type
terms of the form $\lambda e^{\phi}$ may also arise through
divergences, but these will not affect the result). Therefore, our
strategy is this: we will arrange $\Psi$ as an expansion in powers of
$N$, where $N$ is assumed to be large (i.e. of the order of $D \sim
26$). We then look for the contributions order-by-order in $N$ which
contain $\phi\pd_{a}^{2}\phi$. If we work in momentum space, this
involves looking for terms proportional to $p^{2}$. These terms will
be corrections to the coefficient multiplying the Liouville action
(\ref{LA}). (Since we expect $N$ to be of the order of $D$, terms in
(\ref{integral}) of order $1/N,1/N^{2}...$ should be negligible. The
sum of all the diagrams in the expansion will form the exponential of
just the connected diagrams, in the usual way.)

We shall begin by isolating all the terms in (\ref{integral}) which are of
$O(N)$ and have two powers of $\phi$. There turns out to be only one;
this is (writing out the various traces and operators in full)
\bq
O(N) &:& \int\mathcal{D}f\:\exp\left(-\frac{1}{2}\int d^{2}\xi\!\int
d^{2}\xi^{\'}\Delta^{-1}(\xi,\xi^{\'})f(\xi^{\'})\Delta^{-1}(\xi^{\'},\xi)f(\xi)\right)
\nonumber \\
&\times& \frac{1}{2!(4\pi)^{2}}\int d^{2}x_{1}\:\int
d^{2}x_{2}\:f(x_{1})\phi(x_{1})f(x_{2})\phi(x_{2}) \label{ordern}
\eq
We will work in momentum space, and begin by deriving the propagator
associated with the operator
$$\Delta^{-1}(\xi,\xi^{\'})\Delta^{-1}(\xi^{\'},\xi) \equiv
\Gamma^{-1}(\xi,\xi^{\'})$$
This will allow
us to develop a set of rules for representing this and higher-order
integrals as Feynman diagrams. The Fourier transform of the
$\phi$-independent piece of $\Gamma^{-1}$ is given by
\be
I(p)=\frac{1}{(2\pi)^{2}}\int d^{2}q\:\frac{1}{(q^{2}+m^{2})}\frac{1}{((q-p)^{2}+m^{2})}
\ee
We have introduced a small mass $m$ here to regulate the infra-red
divergence of this integral. We find that
\be
I(p) = \frac{1}{4\pi p^{2}(\sqrt{\frac{1}{4}+\frac{m^{2}}{p^{2}}})}\times \ln\left(\frac{\sqrt{\frac{1}{4}+\frac{m^{2}}{p^{2}}}+\frac{1}{2}}{\sqrt{\frac{1}{4}+\frac{m^{2}}{p^{2}}}-\frac{1}{2}}\right)
\ee
which goes like
$$
I(p) \sim \frac{\ln\left(\frac{p^{2}}{m^{2}}\right)}{p^{2}}
$$
as $p/m \- \infty$. Our momentum-space propagator is then the inverse of $I(p)$, which we
will denote by a wiggly line in our Feynman diagram representation: \\ \\
\centerline{$\Gamma
=$\epsfysize=12pt\epsfbox{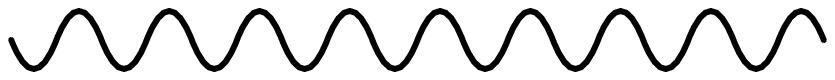}$\sim \frac{p^{2}}{\ln\left(\frac{p^{2}}{m^{2}}\right)}$}
\\ \\
Actually, we can see immediately that the $O(N)$ contribution
(\ref{ordern}) will
be zero, since the diagram is just \\ \\
\centerline{\epsfxsize=5cm\epsfbox{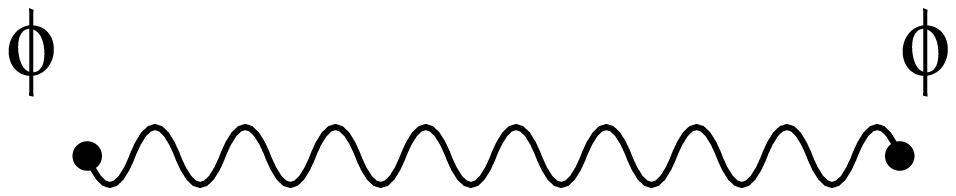}}
\\ \\
which clearly vanishes when we take the regulator $m$ to zero at fixed
momentum $p$.

We now go on to consider the $O(1)$ contributions. We find that there
are four of them; one of which is simply the determinant
$\mbox{Det}^{-\frac{1}{2}}(\Gamma^{-1})$, and the other three can be
represented by the diagrams shown in Figure 1. 

\FIGURE{\centerline{\epsfysize=5cm\epsfbox{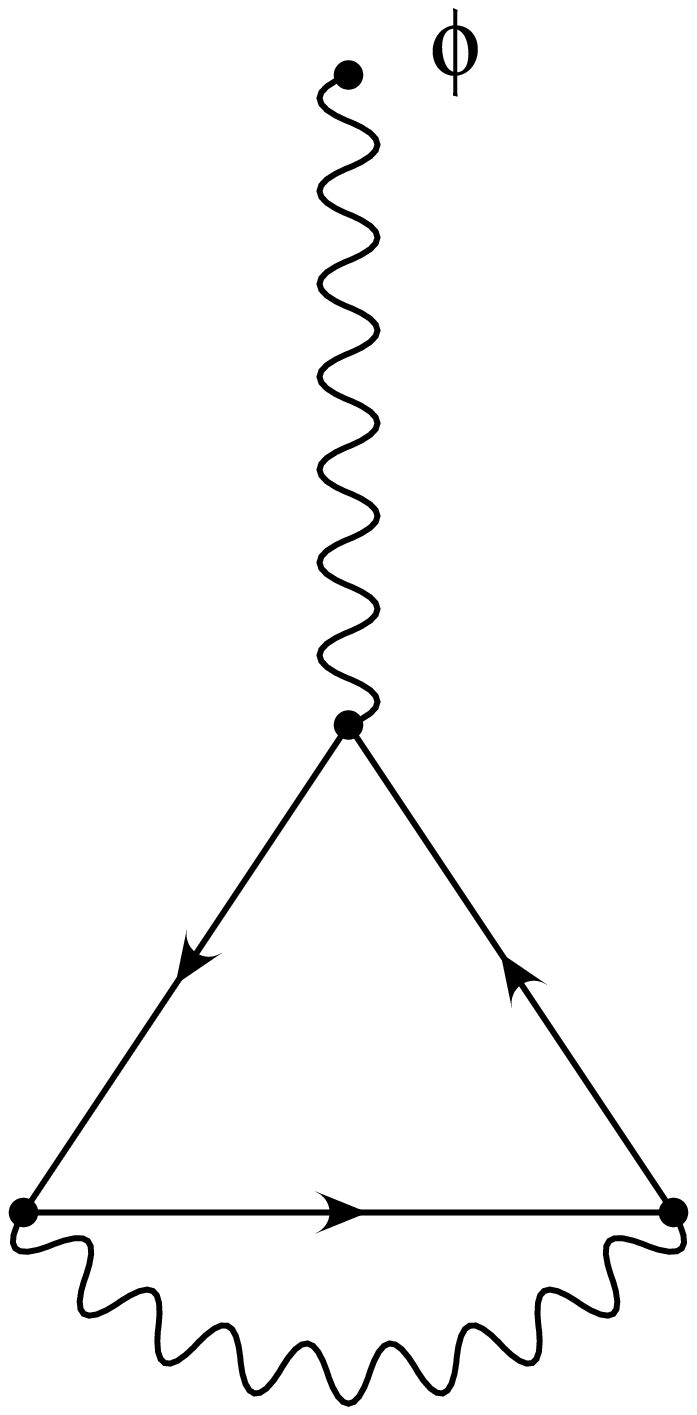}{\hskip
2cm}\epsfysize=5cm\epsfbox{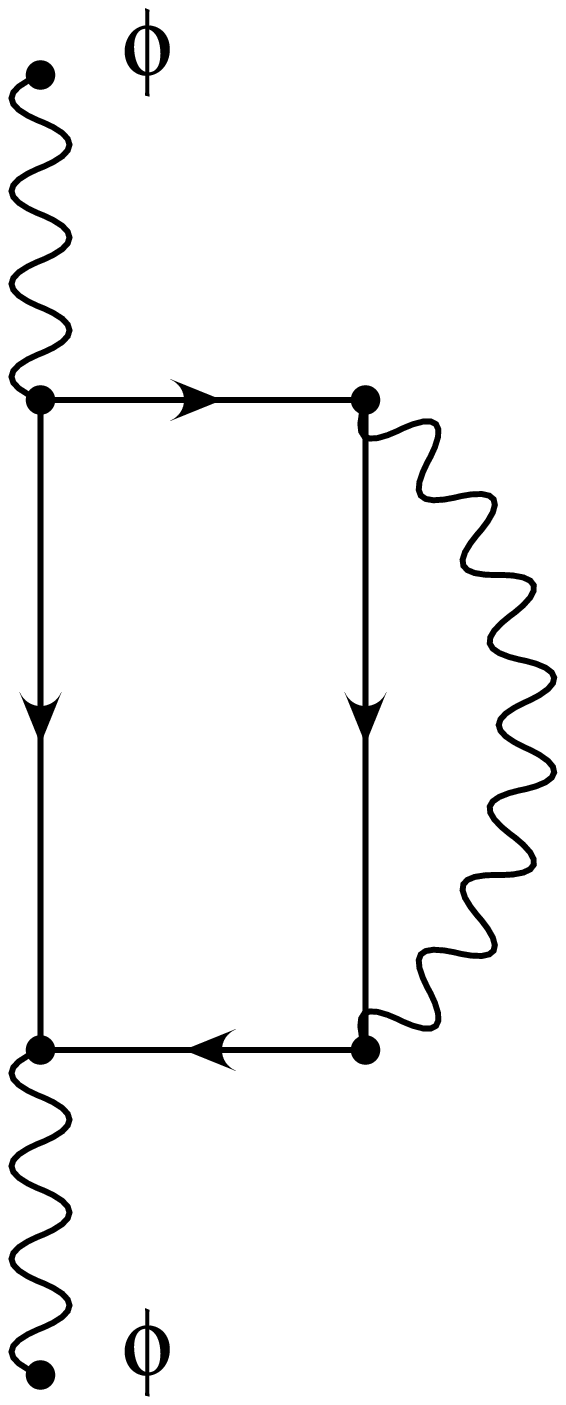}{\hskip
2cm}\epsfysize=5cm\epsfbox{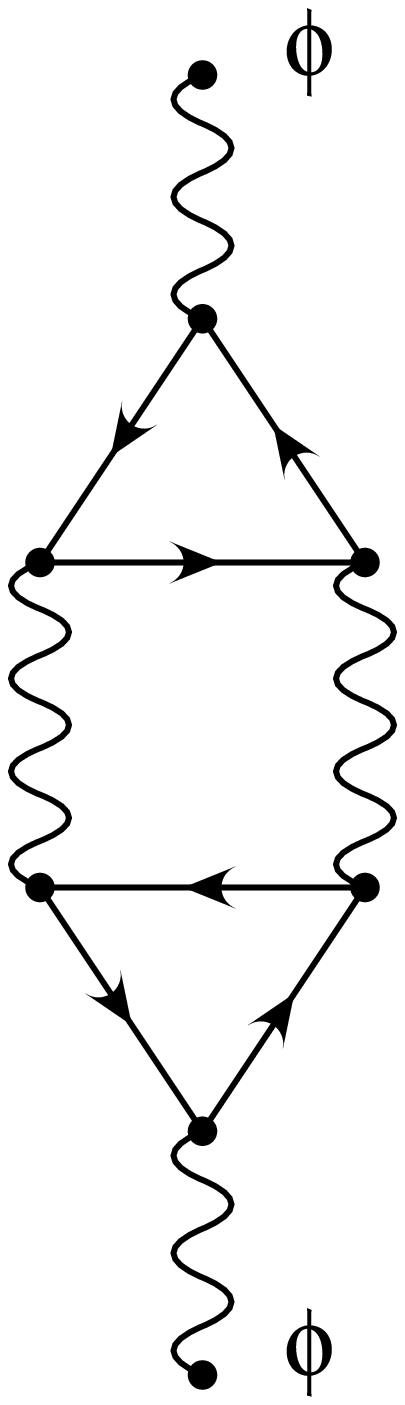}}\caption{The $O(1)$
contributions to (\ref{integral})}}
The solid lines in these diagrams represent the propagator $\Delta^{-1}$. (For instance, the triangles
appearing in the first and third diagrams represent the factors of
$\mbox{Tr}(\Delta^{-1}f)^{3}$ appearing in (\ref{integral})). The first of these diagrams appears to
be proportional only to $\phi$, and would seem to be zero anyway since
it is not 1PI. However, the ``double propagator'' connecting the two
bottom points of the triangle needs to be regulated in a
reparametrisation invariant way, and this will
introduce some extra $\phi$-dependence - we will return
to this issue in a moment. To begin with, we will evaluate the determinant
$\mbox{Det}(\Gamma^{-1})$. The details of this calculation are given
in Appendix B. The result is found to be
$$
\mbox{Det}(\Gamma^{-1}) = \exp\left(\frac{1}{12\pi}\int d^{2}\xi\phi(\xi)\pd_{a}^{2}\phi(\xi)\right)
$$
and hence the correction to the Liouville action factor in $Z$
(equation (\ref{zed})) from this determinant is
\be
\frac{1}{24\pi}\int d^{2}\xi\:(\pd_{a}\phi)^{2}
\ee

We now go on to consider the diagrams in Figure 1. As was mentioned
above, the first diagram requires some extra regularisation, and this
will introduce some extra $\phi$-dependence. We replace the solid line in the double propagator with
$$
G_{\Delta}(x,y) \equiv \int_{\epsilon}^{\infty}ds\:e^{-s\Delta}\delta(x-y)\frac{1}{\sqrt{g(y)}}
$$
and then expand $G_{\Delta}(x,y)$ in powers of $\phi$ about flat
space, as in the determinant calculation above. Analysis of the
integral obtained by using this regulator, along with the other pieces
of the diagram, reveals that there is an overall factor of
$\frac{1}{\ln\epsilon}$ present; this clearly goes to zero when we
send the cutoff to zero, so this diagram does not contribute. 

The second diagram in Figure 1 represents the following asymptotic
(large momentum) integral (up to factors of $\pi$, etc)
$$
\sim \int\!\!d^{2}p\left(\frac{p^{2}}{\ln(\frac{p^{2}}{m^{2}})}\right)^{2}\int\!\!d^{2}j\:d^{2}q\!\frac{1}{(p+q)^{2}+m^{2}}\left(\frac{1}{q^{2}+m^{2}}\right)^{2}\frac{1}{(q+j)^{2}+m^{2}}\frac{j^{2}}{\ln(\frac{j^{2}}{m^{2}})}
$$
The $j$-integral is UV-divergent; we regulate it with a UV cutoff
$\Lambda^{2}$ and subtract the divergence by expanding the
$j$-integral in powers of $q$. What we are left with is proportional
to
$$
\frac{p^{2}}{\ln\left(\frac{p^{2}}{m^{2}}\right)}\times \ln\left(\ln\left(\frac{\Lambda}{m}\right)\right)
$$
and so this diagram also vanishes when we send the $m \rightarrow 0$
at fixed momentum.

We are left with the third diagram of Figure 1 to calculate. This
diagram is in fact proportional to $p^{2}$, and so gives a finite
contribution which we need to calculate. The momentum integral is
rather complicated, but can be evaluated using numerical techniques (a
simple quadrature formula such as Simpson's rule will suffice). We also
need to take into account the various combinatoric factors associated
with the diagram, and the factors arising from the expansion of the
exponential in (\ref{integral}). The total contribution of this
diagram is then found to be approximately
\be
\sim -0.003\times\int d^{2}\xi(\pd_{a}\phi)^{2}
\ee
Hence, at this order the overall coefficient multiplying the Liouville
action in the string partition function is found to be
$$
C = \left(\frac{D-24}{96\pi}\right)+\frac{1}{24\pi}-0.003 \Rightarrow
D_{c} \approx 21
$$
which implies that the critical dimension for $AdS_{D+1}$ space-time is
approximately $D+1 = 22$. In fact, we see a slight departure from the integer
value at this order in $N$ (21.9 to 1 d.p.). Therefore, we might expect that
there ought to be some non-zero contribution at the next order in $N$
which will correct for this. Unfortunately, the $O(1/N)$ diagrams are hard to evaluate; however, we can show that there is some
finite contribution.

Two of the $O(1/N)$ diagrams (see Figure 2) are actually divergent - they are
proportional to $p^{2}\ln(\frac{p^{2}}{m^{2}})$.  Although it is difficult to evaluate either of these diagrams to
check that they exactly cancel, it can be shown quite easily that both are of the form
$$
p^{2}\ln(\frac{p^{2}}{m^{2}})\times \mbox{double integral over
internal momenta}
$$
The internal momentum integrands in each case can be plotted, and are
found to be the same up to an overall factor which we expect to be accounted for
by the combinatorics. A relative minus sign also arises from the
expansion of the exponential; therefore, we expect that these two
divergent diagrams cancel. We are left with three non-zero diagrams at this order, which will
contribute towards the Liouville action coefficient. It is to be hoped
that their contributions are small, and simply tune the result
obtained at $O(1)$ towards the integer value of 22. 
\FIGURE{\centerline{\epsfysize=5cm\epsfbox{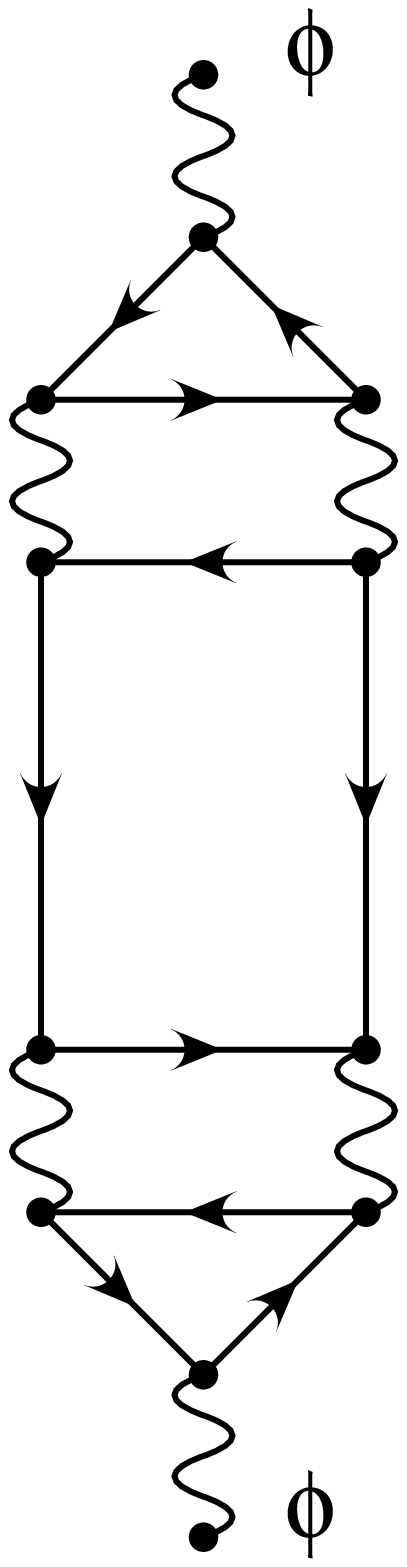}{\hskip
5cm}\epsfysize=5cm\epsfbox{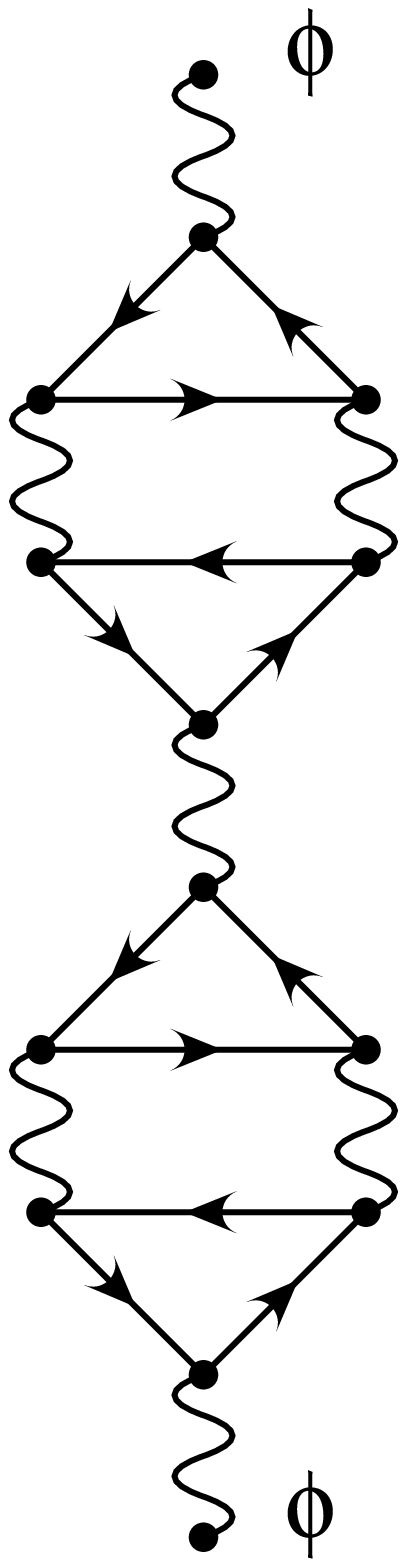}}\caption{The divergent diagrams
at $O(1/N)$}}

\section{Discussion}

Our calculations suggest that the critical dimension for bosonic string theory in
$AdS_{D+1}$ space-time is approximately 22 (that is, an $AdS$ space
comprising 21 $X$-fields and
one $y$-field). This result differs from that derived
in~\cite{deVega}~\cite{27}, where it was found that the critical dimension is 25. Interestingly, this is the
result that would be obtained if all the contributions in our Feynman
diagram expansion cancelled each other. It seems unlikely that this is
the case, since the $O(1)$ contribution is quite large and higher order
terms are expected to be small. There is another reason to suspect
that the critical dimension ought not to be 25. Going back to equation
(\ref{sig}), we can represent the Jacobian $\mbox{Det}^{-1}(\Delta +
f)$ as a functional integration over two worldsheet scalar fields
$v_{1},v_{2}$:
$$
\mbox{Det}^{-1}(\Delta + f) =
\int\mathcal{D}(v_{1},v_{2})\exp\left(-\int
d^{2}\xi\sqrt{g}\:\vec{v}\cdot (\Delta + f)\vec{v}\right)
$$
So we can write $Z$ as
$$
Z=\int \mathcal{D}\psi\mathcal{D}f\:\exp\left(-\int
d^{2}\xi\sqrt{g}\left[f+\vec{\psi}\cdot (\Delta + f)\vec{\psi}\right]\right)
$$
where $\vec{\psi}=(v_{1},v_{2},\vec{W})$ is a $(D+2)$-dimensional vector
of scalar fields for $AdS_{D+1}$. We can now perform the
$f$-integral to obtain a functional delta-function,
$$
Z = \int
\mathcal{D}\psi\:\delta\left[1+\vec{\psi}\cdot\vec{\psi}\right]\exp\left(-\int
d^{2}\xi\sqrt{g}\vec{\psi}\cdot\Delta\vec{\psi}\right)
$$
This is a non-linear sigma model with $O(D+2)$ rotational symmetry. Notice that if the delta functional were not present this would be
the partition function for a bosonic string in $(D+2)$-dimensional flat
space, and we know that the critical dimension of this system is
26. Hence, the critical value of $D$ itself would be 24, implying that
the $AdS$ space from which this came would have a critical dimension
of 25. However, the delta functional depends on the worldsheet
metric! Thus, we cannot simply apply the argument that this is
essentially the partition function for a string in $(D+2)$-dimensional
flat space. We have computed (approximately) the
explicit dependence of this delta functional on the Liouville mode,
and shown it to be non-zero.

\section{Conclusions}

In this paper we have used several different techniques to quantise the Polyakov bosonic string in
$(D+1)$-dimensional $AdS$ space-time. We have shown that the Weyl
structure of the $AdS$ string is essentially the same as the string in
flat space-time; the symmetries of the action obtained after
quantising the $\vec{W}$-fields in these variables dictate that the
partition function must reduce to the Liouville field
theory in the conformal gauge once all the space-time fields have been
integrated out. However, the critical dimension differs
from the flat space-time case; our approximate calculation suggests a
value of 22 for $AdS$, rather
than the value of 26 known for flat space-time.

The fact that we have seen such a significant shift downwards in the
critical dimension is perhaps an encouraging sign that the critical
dimension for the superstring in this space-time may be more
physically 'realistic' than 10. Ideally, we would hope to find a
critical dimension of 5 for the $AdS$ superstring, since this would correspond to a 4-dimensional dual theory in the sense of~\cite{54}. It will therefore be interesting to see the effect of introducing worldsheet supersymmetry into this calculation. (It is worth remembering that this supersymmetric worldsheet theory would still be expected to have a {\it non}-supersymmetric dual gauge theory description.)

In flat space, the vertex operator formalism is used to
calculate string amplitudes. This formalism relies on our ability to
calculate the flat space string partition function. Now that we have
demonstrated the calculation of the $AdS$ string partition function
(at least approximately), it may be that some of the techniques used
here will allow us to begin to compute some amplitudes, and to further
investigate the relationship studied in~\cite{54},~\cite{58} between
this system and the loop equations. With this in mind, the
introduction of a worldsheet boundary (i.e., the extension of this
work to the case of open strings) is essential, since we expect the
ends of an open string to trace out the contour of the dual gauge
theory Wilson loop.

\renewcommand{\theequation}{A-\arabic{equation}}
\setcounter{equation}{0}  
\section{Appendix A: Evaluation of the heat-kernel for $\Omega$}

We begin by writing down the two contributions of the correct order from the second term of (\ref{series}):
\begin{equation}
\int_{0}^{t}dt^{\prime}\int d^{2}\xi^{\prime}K^{t-t^{\prime}}_{\xi\xi^{\prime}}e^{-\phi(\xi)}\left(-\frac{1}{2}(\xi^{\prime}-\xi)^{a}(\xi^{\prime}-\xi)^{b}\left(\partial_{a}\phi(\xi)\partial_{b}\phi(\xi)-\partial_{a}\partial_{b}\phi(\xi)\right)\right)\Delta^{\prime}K^{t^{\prime}}_{\xi^{\prime}\xi}
\end{equation}
and
\begin{equation}
\int_{0}^{t}dt^{\prime}\int d^{2}\xi^{\prime}K^{t-t^{\prime}}_{\xi\xi^{\prime}}e^{-\phi(\xi)}2\left((\xi^{\prime}-\xi)^{a}\partial_{a}\phi(\xi)\partial_{a}\chi(\xi)-(\xi^{\prime}-\xi)^{b}\partial_{b}\partial_{a}\chi(\xi)\right)\partial_{a}^{\prime}K^{t^{\prime}}_{\xi^{\prime}\xi}
\end{equation}
where we have chosen the constant $\bar{\xi} = \xi$. These Gaussian integrals can be evaluated using (\ref{hk}) and are found to equal
\begin{equation}
\frac{1}{24\pi}\left((\partial_{a}\phi(\xi))^{2}-\partial_{a}^{2}\phi(\xi)\right) \label{r1}
\end{equation}
and
\begin{equation}
\frac{1}{4\pi}\left(\partial_{a}^{2}\chi(\xi)-\partial_{a}\phi(\xi)\partial_{a}\chi(\xi)\right) \label{r2}
\end{equation}
respectively.

We now turn to the third term of (\ref{series}). There are four terms
which contribute; the relevant pieces of $V(\xi^{\prime})$ and $V(\xi^{\prime\prime})$ respectively appearing in the integral are
\begin{eqnarray}
(a)&:& -2\partial_{a}\chi(\xi)e^{-\phi(\xi)}\partial_{a}^{\prime} \times  -2\partial_{b}\chi(\xi)e^{-\phi(\xi)}\partial_{b}^{\prime\prime} \nonumber \\
(b)&:& -2\partial_{a}\chi(\xi)e^{-\phi(\xi)}\partial_{a}^{\prime} \times  (\xi^{\prime\prime}-\xi)^{b}\partial_{b}\phi(\xi)e^{-\phi(\xi)}\Delta^{\prime\prime} \nonumber \\
(c)&:& (\xi^{\prime}-\xi)^{a}\partial_{a}\phi(\xi)e^{-\phi(\xi)}\Delta^{\prime} \times  -2\partial_{b}\chi(\xi)e^{-\phi(\xi)}\partial_{b}^{\prime\prime} \nonumber \\
(d)&:& (\xi^{\prime}-\xi)^{a}\partial_{a}\phi(\xi)e^{-\phi(\xi)}\Delta^{\prime} \times (\xi^{\prime\prime}-\xi)^{b}\partial_{b}\phi(\xi)e^{-\phi(\xi)}\Delta^{\prime\prime} \label{4terms}
\end{eqnarray}
These integrals, while being Gaussian, are in practice rather complicated to evaluate. Let us begin with the first of these terms, (a), which involves two single derivatives on the heat-kernels. 
\begin{eqnarray}
(a) &=& \int_{0}^{t}dt^{\prime}\int_{0}^{t^{\prime}}dt^{\prime\prime}\int d^{2}\xi^{\prime}\int d^{2}\xi^{\prime\prime}\frac{e^{\phi(\xi)}}{4\pi(t-t^{\prime})}e^{-\frac{-\vert \xi-\xi^{\prime}\vert^{2}e^{\phi(\xi)}}{4(t-t^{\prime})}} \nonumber \\
&\times& \partial_{a}^{\prime}\left(\frac{e^{\phi(\xi)}}{4\pi(t^{\prime}-t^{\prime\prime})}e^{-\frac{-\vert \xi^{\prime}-\xi^{\prime\prime}\vert^{2}e^{\phi(\xi)}}{4(t^{\prime}-t^{\prime\prime})}}\partial_{b}^{\prime\prime}\left[\frac{e^{\phi(\xi)}}{4\pi t^{\prime\prime}}e^{-\frac{-\vert \xi^{\prime\prime}-\xi\vert^{2}e^{\phi(\xi)}}{4(t^{\prime\prime})}}\right]\right) \nonumber \\
&=&\int_{0}^{t}dt^{\prime}\int_{0}^{t^{\prime}}dt^{\prime\prime}\int d^{2}\xi^{\prime}\int d^{2}\xi^{\prime\prime}\frac{e^{5\phi(\xi)}}{256\pi^{3}}\frac{(\xi^{\prime}-\xi^{\prime\prime})^{a}(\xi^{\prime\prime}-\xi)^{b}}{(t-t^{\prime})(t^{\prime}-t^{\prime\prime})^{2}(t^{\prime\prime})^{2}} \nonumber \\
&\times&e^{e^{\phi(\xi)}(-\frac{\vert \xi-\xi^{\prime}\vert^{2}}{4(t-t^{\prime})}-\frac{\vert \xi^{\prime}-\xi^{\prime\prime}\vert^{2}}{4(t^{\prime}-t^{\prime\prime})}-\frac{\vert \xi^{\prime\prime}-\xi\vert^{2}}{4(t^{\prime\prime})})} \label{bigmomma},
\end{eqnarray}
all multiplied by the factor $4(\partial_{a}\chi)(\partial_{b}\chi)e^{-2\phi(\xi)}$. The simplest way to tackle these integrals is to write them in the form of a matrix equation and use a 'generating function' approach. Begin by making a change of coordinates
$$
X^{a}=(\xi^{'}-\xi^{''})^{a} {\hskip 1cm} Y^{a}=(\xi^{''}-\xi)^{a}
$$
so the $\xi$ part of the integral becomes
$$
\frac{(\partial_{a}\chi)(\partial_{b}\chi)e^{3\phi}}{64\pi^{3}(t-t^{'})(t^{'}-t^{''})^{2}t^{''2}}\int d^{2}X \int d^{2}Y(X)^{a}(Y)^{b} e^{e^{\phi}\left(-\frac{\vert X+Y \vert^{2}}{4(t-t^{\prime})}-\frac{\vert X \vert^{2}}{4(t^{\prime}-t^{\prime\prime})}-\frac{\vert Y \vert^{2}}{4(t^{\prime\prime})}\right)}
$$
We can write the exponent now as \emph{vector $\times$ matrix $\times$ vector},
$$
-\frac{1}{2}\left(X^{1},X^{2},Y^{1},Y^{2}\right){\bf A}\left(X^{1},X^{2},Y^{1},Y^{2}\right)^{T}
$$
where the matrix is
\[ {\bf A} = \frac{e^{\phi}}{2}\left( \begin{array}{cccc}
       \frac{1}{t-t^{'}}+\frac{1}{t^{'}-t^{''}} & 0 & \frac{1}{t-t^{'}} & 0 \\
       0 & \frac{1}{t-t^{'}}+\frac{1}{t^{'}-t^{''}} & 0 & \frac{1}{t-t^{'}} \\
       \frac{1}{t-t^{'}} & 0 & \frac{1}{t-t^{'}}+\frac{1}{t^{''}} & 0 \\
       0 & \frac{1}{t-t^{'}} & 0 & \frac{1}{t-t^{'}}+\frac{1}{t^{''}} \\
\end{array} \right)\]
Let us now define a ``generating function'' $Z(K)$:
$$
Z(K) = \int d^{2}X \int d^{2}Y \exp\left(-\frac{1}{2}(X^{a},Y^{a}){\bf A}(X^{a},Y^{a})^{T}+K(X^{a},Y^{a})^{T}\right)
$$
where $K$ is a vector $(k_{1}^{1},k_{1}^{2},k_{2}^{1},k_{2}^{2})$. This generating function is readily evaluated to be
\begin{equation}
Z(K) = (\mbox{det}{\bf A})^{-1/2}(2\pi)^{2}\exp\left(\frac{1}{2}K{\bf A}^{-1}K^{T}\right) \label{def_f}
\end{equation}
We see that we can now write our integral as
$$
\frac{(\partial_{a}\chi)(\partial_{b}\chi)e^{3\phi}}{64\pi^{3}(t-t^{'})(t^{'}-t^{''})^{2}t^{''2}}(\partial_{1})^{a}(\partial_{2})^{b}Z(K)
$$
evaluated at K=0. Here, $\partial_{1}^{a} \equiv \frac{\partial}{\partial k_{1}^{a}}$. Computing these derivatives and substituting in for $t$ then gives
\begin{eqnarray}
(a) &=& \int_{0}^{t}dt^{'}\int_{0}^{t^{'}}dt^{''}\frac{-\delta^{ab}(\partial_{a}\chi)(\partial_{b}\chi)}{2\pi\sqrt{\frac{t^{2}}{t^{''2}(t^{'}-t^{''})^{2}(t-t^{'})^{2}}}(t-t^{'})(t^{'}-t^{''})tt^{''}} \nonumber \\
&=& -\frac{(\partial_{a}\chi)^{2}}{4\pi} \nonumber
\end{eqnarray}
We now consider term (b) of (\ref{4terms}). In terms of the generating function, this is
$$
(b) = \int_{0}^{t}dt^{'}\int_{0}^{t^{'}}dt^{''}\frac{(\partial_{a}\phi)(\partial_{b}\chi)e^{3\phi}}{64\pi^{3}(t-t^{'})(t^{'}-t^{''})^{2}t^{''}}(\partial_{1})^{a}(\partial_{2})^{b}\left(\frac{1}{t^{''}}-\frac{\partial_{2}^{2}e^{\phi}}{4t^{''2}}\right)Z(K)
$$
It is implicitly understood that this expression is evaluated at $K=0$. This is found to be
\begin{eqnarray}
(b) &=& \int_{0}^{t}dt^{'}\int_{0}^{t^{'}}dt^{''}\frac{\delta^{ab}(\partial_{a}\phi)(\partial_{b}\chi)\left(t-2t^{''}\right)}{2\pi\sqrt{\frac{t^{2}}{t^{''2}(t^{'}-t^{''})^{2}(t-t^{'})^{2}}}(t-t^{'})(t^{'}-t^{''})t^{2}t^{''}} \nonumber \\
&=& \frac{1}{12\pi}(\partial_{a}\phi)(\partial_{a}\chi) \nonumber
\end{eqnarray}
Next, we consider part (c) of (\ref{4terms}). This is
$$
\int_{0}^{t}dt^{'}\int_{0}^{t^{'}}dt^{''}\frac{(\partial_{a}\phi)(\partial_{b}\chi)e^{3\phi}}{64\pi^{3}(t-t^{'})(t^{'}-t^{''})t^{''2}}\left(\partial_{1}+\partial_{2}\right)^{a}(\partial_{2})^{b}\left(\frac{1}{(t^{'}-t^{''})}-\frac{\partial_{1}^{2}e^{\phi}}{4(t^{'}-t^{''})^{2}}\right)Z(K)
$$
again setting $K=0$. This gives
\begin{eqnarray}
(c) &=& \int_{0}^{t}dt^{'}\int_{0}^{t^{'}}dt^{''}\frac{2\delta^{ab}(\partial_{a}\phi)(\partial_{b}\chi)}{2\pi\sqrt{\frac{t^{2}}{t^{''2}(t^{'}-t^{''})^{2}(t-t^{'})^{2}}}t^{2}(t^{'}-t^{''})t^{''}} \nonumber \\
&=& \frac{1}{6\pi}(\partial_{a}\phi)(\partial_{a}\chi) \nonumber
\end{eqnarray}
The last term in (\ref{4terms}), (d), is
$$
\frac{(\partial_{a}\phi)(\partial_{b}\phi)e^{3\phi}}{64\pi^{3} (t-t^{'})(t^{'}-t^{''})t^{''}}(\partial_{1}+\partial_{2})^{a}(\partial_{2})^{b}\left(\frac{1}{t^{'}-t^{''}}-\frac{e^{\phi}\partial_{1}^{2}}{4(t^{'}-t^{''})^{2}}\right)\left(\frac{1}{t^{''}}-\frac{e^{\phi}\partial_{2}^{2}}{4t^{''2}}\right)Z(K),
$$
integrated over $t^{'}$ and $t^{''}$. This gives
\begin{eqnarray}
(d) &=& -\int_{0}^{t}dt^{'}\int_{0}^{t^{'}}dt^{''}\frac{\delta^{ab}(\partial_{a}\phi)(\partial_{b}\phi)\left(t^{2}-2tt^{'}+6t^{'}t^{''}-6t^{''2}\right)}{2\pi\sqrt{\frac{t^{2}}{t^{''2}(t^{'}-t^{''})^{2}(t-t^{'})^{2}}}t^{3}(t-t^{'})(t^{'}-t^{''})t^{''}} \nonumber \\
&=& -\frac{(\partial_{a}\phi)^{2}}{24\pi} \nonumber
\end{eqnarray}

And so finally, combining results $(a), (b), (c)$ and $(d)$ with
(\ref{r1}) and (\ref{r2}) we find the heat-kernel for the operator $\Omega$ to be
\begin{equation}
K(\xi,\xi;\epsilon) = e^{2\chi(\xi)-\phi(\xi)}\left[\frac{e^{\phi}}{4\pi\epsilon}-\frac{1}{24\pi}\partial_{a}^{2}\phi+\frac{1}{4\pi}\left(\partial_{a}^{2}\chi-(\partial_{a}\chi)^{2}\right)+O(\epsilon)\right]
\end{equation}

\renewcommand{\theequation}{B-\arabic{equation}}
\setcounter{equation}{0}  
\section{Appendix B: Evaluation of $\mbox{Det}(\Gamma^{-1})$}

To calculate this determinant, we begin by noting that 
$$
\delta_{\phi}\Delta^{-1} =
-\Delta^{-1}(\delta_{\phi}\Delta)\Delta^{-1} = \Delta^{-1}(\delta\phi)
$$
and so 
\be
\delta_{\phi}\left(\Delta^{-1}(\xi,\xi^{\'})\Delta^{-1}(\xi,\xi^{\'})\right)
= 2\left(\Delta^{-1}(\xi,\xi^{\'})\Delta^{-1}(\xi,\xi^{\'})\right)(\delta\phi(\xi^{\'}))
\ee
which implies that
$$
\delta_{\phi}\Gamma = -2(\delta\phi)\Gamma
$$
Now,
$\delta_{\phi}\ln\mbox{Det}(\Gamma^{-1})=-\delta_{\phi}\ln\mbox{Det}(\Gamma)$,
so we can use an identity analogous to (\ref{rep}) to obtain
$$
\delta_{\phi}\ln\mbox{Det}(\Gamma^{-1}) = 2\mbox{Tr}(\delta\phi)e^{-\epsilon\Gamma}
$$
We now proceed as before, making an expansion in powers of $\phi$
about the $\phi$-independent piece of $\Gamma$. If we work in momentum
space we are led to consider the following expression:
\be
\mbox{Tr}(\delta\phi)e^{-\epsilon\Gamma}=\frac{2}{(2\pi)^{4}}\int_{0}^{\epsilon}d\tau\!\!\int
d^{2}q\:d^{2}p\:\delta\phi(-p)\phi(p)e^{-(\epsilon-\tau)\tilde{\Gamma}(p+q)}\frac{\pd}{\pd\tau}e^{-\tau\tilde{\Gamma}(p)} 
\ee
where we are using the asymptotic expression for the Fourier transform
of $\Gamma$, so
$$
\tilde{\Gamma}(p) \equiv \frac{p^{2}}{\ln\left(\frac{p^{2}}{m^{2}}\right)}
$$
We can see that to obtain a non-zero result
as we send the regulator $\epsilon \rightarrow 0$, we need a factor of
$1/ \epsilon^{c}$ (where $c$ is some number) to be generated from the
integral over $q$. Therefore, we need to study the large-$q$ regime of
this integral. We thus proceed by making an expansion for $p << q$, and examining the terms
proportional to $p^{2}$. We make a change of variables such that
$Q=q^{2}$, and then put $x=Q / \ln Q$. Our integral is now
$$
-\frac{2}{(2\pi)^{4}}\:\pi
p^{2}\int_{0}^{\epsilon}d\tau\int_{0}^{\infty}dx\:\frac{dQ}{dx}\:x\:e^{-\epsilon
x} \left((\epsilon-\tau)\tilde{f}_{1}(Q)+(\epsilon-\tau)^{2}\tilde{f}_{2}(Q)\right),
$$
integrated over $p$. The functions $\tilde{f}_{1}$ and $\tilde{f}_{2}$ are the Taylor coefficients arising from the
expansion in powers of $p$. Integrating out $\tau$ then leaves us with an
integral over $x$ of the form
$$
\int_{0}^{\infty}dx\:e^{-\epsilon x}\left(\epsilon^{2}F_{1}(x)+\epsilon^{3}F_{2}(x)\right)
$$
($F_{1}(x)$ and $F_{2}(x)$ are simply the Taylor coefficients,
re-expressed in terms of $x$). Now, if we replace the
lower limit in this integral with some number $a$, we know that the
result should be independent of $a$ as we send $\epsilon \rightarrow
0$. Thus, the integral can be computed numerically and is found to be
approximately $\pi/6$. In fact, comparison with
the corresponding expression for the operator $\Delta$ (which we can
evaluate exactly, as in~\cite{13}) shows that the two numerical
integrals obtained with this technique are the same in both cases, and therefore we know that the
{\it exact} result must be

\be
\delta_{\phi}\ln\mbox{Det}(\Gamma^{-1}) = \frac{1}{6\pi}\int d^{2}\xi
\delta\phi(\xi)\pd_{a}^{2}\phi(\xi)
\ee

\acknowledgments

Ian Davies would like to thank James Gregory, Ken Lovis, Tony Padilla and
David Page for helpful discussions. Ian Davies is supported by a PPARC
research studentship. 

\bibliographystyle{jhep}  
  
\bibliography{ref}

\end{document}